\documentclass[aip,jmp, amsmath,amssymb, preprint]{revtex4-1}

\usepackage{graphicx}% Include figure files
\usepackage{dcolumn}% Align table columns on decimal point
\usepackage{bm}% bold math
\usepackage{mathtools}
\usepackage{amsmath}
\usepackage{amssymb}
\usepackage{amsthm}
\usepackage{mathrsfs}
\usepackage{yfonts}
\usepackage{txfonts}

\newcommand{\M}[0]{\ensuremath{\mathbf{M}}}
\newcommand{\T}[0]{\ensuremath{\mathrm{T}}}

\newcommand{\GLs}[0]{\ensuremath{\GL^{+3}}}
\newcommand{\GGG}[0]{\ensuremath{\mathbf{G}}}

\newcommand{\q}[0]{\ensuremath{\mathbf{q}}}

\newcommand{\vv}[0]{\ensuremath{\mathbf{v}}}

\newcommand{\QQ}{\mathbf{Q}}
\newcommand{\WW}{\mathbf{W}}
\newcommand{\SSS}{\mathbf{S}}
\newcommand{\R}[0]{\ensuremath{\mathbb{R}}}
\newcommand{\N}[0]{\ensuremath{\varmathbb{N}}}
\newcommand{\D}[0]{\ensuremath{\mathcal{D}}}
\newcommand{\HD}[0]{\ensuremath{\hat{\mathcal{D}}}}
\newcommand{\dd}[0]{\ensuremath{\mathrm{d}}}
\newcommand{\hd}[0]{\ensuremath{\hat{\mathrm{d}}}}

\newcommand{\ee}[0]{\ensuremath{\mathbf{e}}}
\newcommand{\EE}[0]{\ensuremath{\mathbf{E}}}
\newcommand{\AAA}[0]{\ensuremath{\mathbf{A}}}
\newcommand{\HAA}[0]{\ensuremath{\hat{\mathbf{A}}}}

\newcommand{\HA}[0]{\ensuremath{\hat{\mathbf{A}}}}
\newcommand{\RR}[0]{\ensuremath{\mathbf{R}}}
\newcommand{\HR}[0]{\ensuremath{\hat{\mathbf{R}}}}
\newcommand{\TT}[0]{\ensuremath{\mathbf{T}}}
\newcommand{\KK}[0]{\ensuremath{\mathbf{K}}}

\newcommand{\SO}[0]{\ensuremath{\mathbf{SO}}}

\newcommand{\GL}[0]{\ensuremath{\mathbf{GL}}}

\newcommand{\Td}[0]{\ensuremath{\mathbb{T}}}

\newcommand{\hs}[0]{\hspace*{5.9mm}}

\newcommand{\bleps}[0]{\ensuremath{\boldsymbol{\varepsilon}}}

\newcommand{\BSigma}[0]{\ensuremath{\boldsymbol{\Sigma}}}

\newcommand{\blpi}[0]{\ensuremath{\boldsymbol{\pi}}}

\newcommand{\PP}[0]{\ensuremath{\mathbf{P}}}

\newcommand{\xxx}[0]{\ensuremath{\mathbf{x}}}
\newcommand{\y}[0]{\ensuremath{\mathbf{y}}}

\newcommand{\HH}[0]{\ensuremath{\mathbf{H}}}
\newcommand{\FFF}[0]{\ensuremath{\mathbf{F}}}

\newcommand{\vb}[1]{\ensuremath{\big|_{#1}}}

\newcommand{\ints}{\int\limits_{\Sigma}}
\newcommand{\Hil}{\mathscr{H}}
\newcommand{\eul}{\text{e}}
\newcommand{\ham}{\mathsf{H}}
\newcommand{\Lfun}[3]{\mathsf{L}^{#1}\left(#2,#3\right)}

\newcommand{\im}{\mathsf{i}}
\newcommand{\XXX}{\mathfrak{X}}
\newcommand{\Uni}{\mathsf{U}}
\newcommand{\Apx}{\mathsf{A}_{\xxx}}
\newcommand{\Aps}{\mathsf{A}_{\Sigma}}
\newcommand{\UL}{\Uni^{\text{L}}}

\newcommand{\Upi}{\Uni^{\piup}}
\newcommand{\Lgen}{\mathsf{L}}
\newcommand{\Llgen}{\mathsf{L}^{(\lambda)}}
\newcommand{\LEgen}{\mathsf{L}^{(\EE)}}
\newcommand{\LLlgen}{L^{(\lambda)}}
\newcommand{\LLEgen}{L^{(\EE)}}
\newcommand{\bldel}{\mathbf{\Delta}}
\newcommand{\kk}{\mathbf{k}}
\newcommand{\hh}{\mathbf{h}}

\newcommand{\ems}{\mathfrak{e}}
\newcommand{\EQC}{\mathfrak{C}(\Hil_{\Sigma})}
\newcommand{\conf}{\mathfrak{conf}}
\newcommand{\Conf}{\mathfrak{Conf}}

\begin{document}
%\preprint{AIP/123-QED}
% Use the \preprint command to place your local institutional report number 
% on the title page in preprint mode.
% Multiple \preprint commands are allowed.

\title[]{Kinematical Hilbert Space for Einstein-Cartan Theory} %Title of paper

% repeat the \author .. \affiliation  etc. as needed
% \email, \thanks, \homepage, \altaffiliation all apply to the current author.
% Explanatory text should go in the []'s, 
% actual e-mail address or url should go in the {}'s for \email and \homepage.
% Please use the appropriate macro for the type of information

% \affiliation command applies to all authors since the last \affiliation command. 
% The \affiliation command should follow the other information.
\author{Mari\' an Pilc}
\email[]{marian.pilc@gmail.com}
\noaffiliation
\affiliation{Institute of Theoretical Physics,Faculty of Mathematics and Physics, Charles University,V Holesovickach 2,
180 00 Praha 8, Czech Republic}
%\homepage[]{Your web page}
%\thanks{}
%\altaffiliation{}
% Collaboration name, if desired (requires use of superscriptaddress option in \documentclass). 
%\noaffiliation is required (may also be used with the \author command).
%\collaboration{}
%\noaffiliation

\date{\today}

\begin{abstract}
Kinematical Hilbert space for Einstein-Cartan theory is constructed via von Neumann ideas of infinity-dimensional tensor product of Hilbert spaces. Field of comframe is considered as basic variable what is in contrast with standard euclidean LQG which is build by Wilson loops of Ashtekar-Barbero-Immirzi connection. 
\end{abstract}

%\pacs{123456}% insert suggested PACS numbers in braces on next line
\maketitle %\maketitle must follow title, authors, abstract and \pacs

%----------------------------------------------------
%						Introduction
%----------------------------------------------------	
\section{Introduction}
In a modern theory of quantum gravity (LQG) the Wilson loops play crucial role in the construction of the auxiliar Hilbert space. In the euclidean version kinematical space is given by $\Hil_{\text{LQG}}=L^2(\XXX,\dd\mu)$, where $\XXX$ is the space of classes of Ashtekar-Barbero-Immirzi connection up to gauge transformation of diffeomosphisms and local $\SO(3)$. Since the LQG works with partially solved gauge freedom only partial Dirac observables can be quantized on the space $\Hil_{\text{LQG}}$. This may produce several, perhaps hypothetical, problems like absence of the crucial geometrical and physically measurable objects like metric tensor, (co)frames, or curvature in a final picture. Of course they can not be represented on the physical Hilbert space given by solution of all constrains,
but the question is whether such representation exists and if the answer is affirmative how it is related to the standard Lorentzian\cite{Livine} loop approach. The construction of such representation is the key point of this article, the question of the relation with standard LQG is kept for future research at this moment, but if there is any relation with LQG, then one may expect that it should be found after solving spatial diffeomorphism plus Lorentz constrains.\\
\hs The article is organizes as follows. In the section \ref{phase_space} the results of previous works\cite{Nikolic}$^,$\cite{Pilc} are summarized. The point version of the phase space is quantized in the section \ref{Point_algebra}. These results are used with help of ideas of von Neumann construction of infinity-dimensional tensorial product (summarized in section \ref{Tensor_prd}) in construction of wanted representation in section \ref{Quantum_algebra}.

%----------------------------------------------------
%						Phase space od ECT
%----------------------------------------------------	

\section{Phase space of Einstein-Cartan Theory}\label{phase_space}
Einstein-Cartan theory is a gauge theory where local Poincar\' e 
group plays a role of gauge symmetry\cite{Kibble}$^,$\cite{Hehl1}$^,$\cite{Hehl2}. Full configuration space
of Einstein-Cartan theory is given by orthonormal coframe $\ee^a$ ($a,b,\ldots=0,1,2,3$) and metric-compatible connection  $\HAA^{ab}=-\HAA^{ba}$. The hat over $\HAA^{ab}$ means that associated covariant derivative operator $\HD$,
is acting on the spacetime manifold $\M$. Similar for $\hd$. "Hat-free" objects $\AAA^{ab}$, $\dd$, $\D$ are reserved for objects acting on the spatial section $\Sigma$. We are assuming that $\Sigma$ is compact orientable manifold, e.g. torus, and variables $\ee^a$ and $\HA^{ab}$ are globally defined. The second assumption is motivated by the Geroch's condition\cite{Geroch} which guarantees the existence of the global spinor structure over the manifolf $\M$.  Let $\HR^a_{\phantom{a}b}$ be the curvature of the connection $\HD$ then action of the Einstein-Cartan theory can be written in the form 
\begin{eqnarray}
S=\int\limits_{\M}-\frac{1}{32\pi\kappa}\bleps_{abcd}
\eta^{b\bar{b}}\hat{\RR}^{a}_{\phantom{a}\bar{b}}\wedge\ee^c\wedge\ee^d,\label{Cartan_action}
\end{eqnarray}
where $\kappa$ is Newton's constant (c=1). We are using spatial negative signature of the metric, i.e.
$(\eta_{ab})=\text{diag}(+1,-1,-1,-1)$. Equations of the motion given by the action (\ref{Cartan_action})
are
\begin{eqnarray}
0&=&\,\,\,\,\,\frac{1}{8\pi\kappa}\bleps_{abcd}\ee^c\wedge\HD\ee^d\,\,=
-\frac{1}{8\pi\kappa}\left(\hat{T}^c_{ab}+\hat{T}^d_{da}\delta^c_b-\hat{T}^d_{db}\delta^c_a\right)\hat{\BSigma}_c,\label{TorEQ}\\
0&=&-\frac{1}{16\pi\kappa}\bleps_{abcd}\HR^{bc}\wedge\ee^d=-\frac{1}{8\pi\kappa}\hat{G}^c_{\phantom{c}a}\hat{\BSigma}_c,\label{EEQ}
\end{eqnarray}
where the torsion components are given by
\begin{eqnarray}
\HD\ee^a=\hat{\TT}^a=\frac{1}{2}\hat{T}^a_{bc}\ee^b\wedge\ee^c,\nonumber
\end{eqnarray}
3-volume forms
\begin{eqnarray}
\hat{\BSigma}_a=\frac{1}{3!}\bleps_{abcd}\ee^{b}\wedge\ee^{c}\wedge\ee^d\nonumber,
\end{eqnarray}
and $\hat{G}^a_{\phantom{a}b}$ is the Einstein tensor
\begin{eqnarray}
\hat{G}^a_{\phantom{a}b}&=&\hat{R}^{ca}_{\phantom{ab}cb}-\frac{1}{2}\hat{R}^{cd}_{\phantom{cd}cd}\delta^a_b,\nonumber\\
\HR^{ab}&=&\frac{1}{2}\hat{R}^{ab}_{\phantom{ab}cd}\ee^c\wedge\ee^d.\nonumber
\end{eqnarray}
Equation (\ref{TorEQ}) implies that connection $\HD$ is torsion-free and together with 
metricity of $\HD$ we have that $\HD$ is geometrical connection. Equations (\ref{EEQ}) are Einstein equations of General Relativity.\\
\hs Let us summarize the results given by previous work\cite{Pilc} for the Dirac-Hamiltonian formulation.
\nobreak{(3+1)-decomposition} of basic variables are given by expressions ($\alpha,\beta,\ldots=1,2,3$ are spatial coordinate indices)
\begin{eqnarray}
&\ee^a&=\lambda^a\dd t+\EE^a=\lambda^a\dd t+ E^{a}_{\alpha}\dd x^{\alpha},\label{ee^a}\\
&\HAA^{ab}&=\Lambda^{ab}\dd t+\AAA^{ab}.\label{AAA^{ab}}
\end{eqnarray}
It is useful for our purposes to decompose even the vector frame $\ee_a$ into spatial and time parts
\begin{eqnarray}
\ee_a=\lambda_a\partial_t+\EE_a=\lambda_a\partial_t+E_a^{\alpha}\partial_{\alpha}.\label{ee_a}
\end{eqnarray}
It should be noted that $\lambda_a\neq\eta_{ab}\lambda^a$. We hope that this notation is not 
confusing since if we need to in$/$de-crease indices then it will be explicitly written 
using metric tensor. Variables $\lambda^a$, $\lambda_a$, $\EE^a$ and $\EE_a$ are not independent and we can 
express vector coefficients by using the covectors via well known formula for inverse matrix
\begin{eqnarray}
e\lambda_a&=&\frac{\partial \,e\,\,}{\partial\lambda^a},\\
e E^{\alpha}_a&=&\frac{\partial \,e\,\,}{\partial E^a_{\alpha}},
\end{eqnarray}
where 
\begin{eqnarray}
e=\frac{1}{3!}\bleps_{abcd}\bar{\varepsilon}^{\alpha\beta\gamma}\lambda^aE^b_{\alpha}E^c_{\beta}E^d_{\gamma}
\end{eqnarray}
is determinant of matrix $(\lambda^a,E^a_{\alpha})$. It is easy to see that $\lambda^a\lambda_b$ or
$\EE^a_b=\EE^a(\EE_b)$ are projections to time or spatial subspaces of $\Td_{\Sigma}\M$, respectively.
\\
\hs The Hamiltonian $\ham$ is given by sum of the first class constraints
\begin{eqnarray}
&\blpi(N)&=\int\limits_{\Sigma}\blpi_aN^a,\label{pi(N)}\\
&\RR(\mu)&=\int\limits_{\Sigma}\frac{1}{16\pi\kappa}\bleps_{abcd}\mu^a\RR^{bc}
					\wedge\EE^d=\int\limits_{\Sigma}\mu^a\RR_a,\label{RR(mu)}\\
&\TT(\Lambda)&=\int\limits_{\Sigma}-\frac{1}{32\pi\kappa}\bleps_{abcd}\D\Lambda^{ab}
					\wedge\EE^c\wedge\EE^d=
\int\limits_{\Sigma}-\frac{1}{16\pi\kappa}\bleps_{abcd}\Lambda^{ab}\wedge\EE^c\wedge\D\EE^d=
\int\limits_{\Sigma}\frac{1}{2}\Lambda^{ab}\TT_{ab},\label{TT(Lambda)}
\end{eqnarray}
where $N^a$, $\mu^a$, $\Lambda^{ab}$ play role of Lagrange multipliers, $\blpi_a$ is a conjugate momentum to $\lambda^a$, $\D$ is $\SO(\eta)$ connecion over $\Sigma$ defined for all vector-forms $\vv^a$ via
\begin{eqnarray}
\D\vv^a=\dd\vv^a+\AAA^{ab}\eta_{bc}\wedge\vv^a
\end{eqnarray}
and $\RR^{ab}=\dd\AAA^{ab}+\eta_{cd}\AAA^{ac}\wedge\AAA^{db}$ is a curvature of $\D$.\\
\hs Symplectic structure is given by Dirac brackets
\begin{eqnarray}
\begin{matrix}
\{\,\EE(\QQ),\GGG(\KK)\,\}^*&=&\ints\QQ_a\wedge\KK^a,\\ 
\{\,\WW(\lambda)\,,\,\blpi(N)\,\}^*&=&\ints\WW_aN^a\\
\end{matrix}
\label{sympl_str}
\end{eqnarray}
where only the non-trivial brackets are explicitly written. We have used smeared variables in (\ref{sympl_str}) given by
\begin{eqnarray}
&\EE(\QQ)&=\ints\QQ_a\wedge\EE^a,\nonumber\\
&\GGG(\KK)&=\ints\GGG_a\wedge\KK^a=\ints-\frac{1}{16\pi\kappa}\bleps_{abcd}\KK^a\wedge\AAA^{bc}\wedge\EE^d\nonumber\\
&\WW(\lambda)&=\ints\WW_a\lambda^a,\nonumber
\end{eqnarray}
where $\QQ_a$, $\KK^a$, $\WW_a$ are smearing forms and $\GGG_a$ is a canonical momentum conjugated to $\EE^a$. Since $\GGG_a$ has only twelve degrees of freedom(DOFs) per a point the rest of $\AAA^{ab}$ with eighteen DOFs per point should be established by the second class constrain given by Hamilton-Dirac procedure
\begin{eqnarray}
&\SSS^{ab}&=\frac{1}{8\pi\kappa}\EE^{(a}_{\phantom{c}}\EE^{b)}_c\wedge\D\EE^c=0.
\end{eqnarray}
In addition let us assume that the orthonormal coframe $\ee^a$ is future and righthand oriented, then the configuration space is given by the infinity-dimensional manifold
\begin{eqnarray}
\Conf=\{(\lambda^a,\EE^a);\, e>0,\, \eta_{ab}\lambda^a\lambda^b>0,\, \lambda^0>0, \, 
\q<0\},\nonumber
\end{eqnarray}
where $\q=\eta_{ab}\EE^a\otimes\EE^b$ is a spatial metric tensor. The cotangent bundle $\Td^*\Conf$ forms our phase space with symplectic structure given by (\ref{sympl_str}).\\
\hs Here our overview of classical results is finished and we can start to build the quantum formulation.

%-----------------------------------------------------------------------------------------
%								Quantum Preliminaries
%-----------------------------------------------------------------------------------------

\section{Quantum Preliminaries}\label{quant_prelimi}

Before we start to construct Hilbert space of Einstein-Cartan theory let us focus our attention to the following simple excersice well known from the quantum mechanics of the particle moving on the half line. Canonical variables of this system are $x$ and $p$, where
$x$ is a position of the particle on the half line $x>0$ and $p$ is its canonical momentum. We can naively
represent them  on $\Hil=L^2(\R^+,\dd x)$ as $\varrho(x)=x$, $\varrho(p)=-\im\partial_x$. The operators 
$\varrho(x)$ and $\varrho(p)$ are symmetric but $\varrho(p)$ can not be extended into selfadjoint operator 
on $\Hil$. In order to see this let us compute its deficiency indices $n_{\varepsilon}$, where 
$\varepsilon=\pm 1$. Equations
\begin{eqnarray}
-\im\partial_x\psi^{(\varepsilon)}-\im\varepsilon\psi^{(\varepsilon)}=0 \nonumber
\end{eqnarray}
have solutions
\begin{eqnarray}
\psi^{(\varepsilon)}=A^{(\varepsilon)}\eul^{-\varepsilon x}.\nonumber
\end{eqnarray}
Solution $\psi^{(+1)}$ belongs to the space $L^2(\R^+,\dd x)$ while $\psi^{(-1)}$ is not square 
integrable function on $\R^+$. Since $n_+=1$ and $n_-=0$ we have $n_+\neq n_-$. Thus we can not construct 
the selfadjoint extenstion of the operator $-i\partial_x$. Hence if one wants to describe the quantum 
particle on the half line then one has to choose different set of basic variables. The first observation 
is that $\R^+$ is a group $\GL^+(\R)$. Invariant measure on $\GL^+(\R)$ is $\omega_{\GL^+(\R)}=
\frac{\dd x}{x}$ hence the good candidate for the "momentum" operator is given 
by $\varrho(xp)=-\im x\partial_x$. Indeed, the operator $\varrho(xp)$ is symmetric on $L^2(\R^+,\frac{\dd x}{x})$.
\begin{eqnarray}
\langle\psi_2|\varrho(xp)\psi_1\rangle=\int\limits_{\R^+}\frac{\dd x}{x}\overline{(\psi_2)}(-\im x\partial_x\psi_1)=
\int\limits_{\R^+}\frac{\dd x}{x}\overline{(-\im x\partial_x\psi_2)}\psi_1=\langle\varrho(xp)\psi_2|\psi_1\rangle\nonumber
\end{eqnarray}
and its deficiency indices are determined by the following equations
\begin{eqnarray}
-\im x\partial_x\psi^{(\varepsilon)}-\im\varepsilon\psi^{(\varepsilon)}=0 \nonumber
\end{eqnarray}
with solutions
\begin{eqnarray}
\psi^{(\varepsilon)}=A^{(\varepsilon)}x^{-\varepsilon},\nonumber
\end{eqnarray}
which do not belong to $L^2(\R^+,\frac{\dd x}{x})$ if $A^{(\varepsilon)}\ne 0$. Hence $n_+=n_-=0$ and the operator $\varrho(xp)$ is 
essentially selfadjoint. The algebra of the basic variables is a space spanned on operators $\varrho(x), 
\varrho(xp)$ with nontrivial commutator
\begin{eqnarray}
[\varrho(x),\varrho(xp)]=\im\varrho\left(\{x,xp\}\right)=\im\varrho(x)\nonumber.
\end{eqnarray}
\hs As we have seen on this simple exercise the choice of the basic variables plays the crucial role in 
the context of quantization. In the next section we will try to understand a point version of
Einstein-Cartan phase space. 

%-----------------------------------------------------------------------------------------------
%			Point Algebra of Basic Variables
%-----------------------------------------------------------------------------------------------

\section{Point Algebra of Basic Variables}\label{Point_algebra}
Let us focus in this section on the introduction of a Hilbert space $\Hil_{\xxx}$ associated with an
arbitrary point $\xxx$ in the spatial section $\Sigma$. We will define a point representation of the 
basic variables related to the canonical coordinates on the phase space $\Td^*\Conf$. Let us mention 
that all canonical variables $\lambda^a(\xxx)$, $\EE^{a}(\xxx)$, $\blpi_a(\xxx)$, 
$\GGG_a(\xxx)$ are local functions of the point $\xxx$. No derivatives, no complicated integrals or any kind of dislocation are presented, hence we can explore them in the single point $\xxx$. Before we start, we will introduce  spacetime notation\footnote{Explicit writing of the point $\xxx$ is omitted till the end of this
section.}
\begin{eqnarray}
e^a_{\mu}=(e^a_t=\lambda^a;e^a_{\alpha}=E^a_{\alpha}),\nonumber\\
p_a^{\mu}=(p_a^t=\pi_a;p_a^{\alpha}=G^{\alpha}_a),\nonumber
\end{eqnarray}
Point version of cannonical momenta are given by 
\begin{eqnarray*}
\tilde{\pi}_a(\xxx ')={\pi}_a\delta_{\xxx\xxx '},\\
\tilde{G}^{\alpha}_a(\xxx ')=G^{\alpha}_a\delta_{\xxx\xxx '},
\end{eqnarray*}
where $\blpi_a=\tilde{\pi}_a\dd^3x$ and $\GGG_a=\frac{1}{2}\tilde{G}^{\alpha}_a\varepsilon_{\alpha\beta\gamma}\dd x^{\beta}\wedge\dd x^{\gamma}$.
Since we are working with the point variables, their canonical relations are given by
\begin{eqnarray}
\{e^a_{\mu},p^{\nu}_b\}=\delta^a_b\delta^{\nu}_{\mu}\nonumber
\end{eqnarray}
and the phase space is defined in accordance to the Einstein-Cartan phase space as $\Td^*\conf$, where
\begin{eqnarray}
\conf=\{(e^a_{\mu}); e=\det(e^a_{\mu})>0,\,\eta_{ab}e^a_te^b_t>0,\,e^0_t>0,\,
\eta_{ab}e^a_{\alpha}e^b_{\beta}<0 \}.\nonumber
\end{eqnarray}
Thanks to the positivity of the determinant $e$ we can see that $\conf\subset\GL^+(\R^4)\equiv\GL^+$, anyway the subset
$\conf$ is not a group. Now we will try to construct a representation of the basic variables. Let us define
a Hilbert space $\Hil_{\xxx}\equiv\Hil$ as a space of square integrable functions over $\conf$ 
\begin{eqnarray}
\Hil=\Lfun{2}{\conf}{\frac{\dd e}{e^4}},
\end{eqnarray}
where $\frac{\dd e}{e^4}$ is left/right-invariant\footnote{In the case of the general noncompact group it may 
happen that left and right invariant measures are not equal.} Haar's measure on the $\GL^+$, which is unique 
up to the multiplicative constant. $\dd e=\dd e^0_t\dd e^0_x\dots\dd e^3_y\dd e^3_z$ is Lebesgue measure 
on the coordinates $(e^a_{\mu})\in\R^{4\times 4}$ of the space $\conf$.\
The representation $\varrho$ of $e^a_{\mu}$ is given by trivial multiplication 
\begin{eqnarray}
\varrho(e^a_{\mu})\psi(e^a_{\mu})=e^a_{\mu}\psi(e^a_{\mu}).\nonumber
\end{eqnarray}
It is well known fact that such operators can be extended into the selfadjoint operators. 
The problems occure with variables $p^a_{\mu}$, 
since the action of $\varrho(p^a_{\mu})=-\im\partial_{e^a_{\mu}}$ given by the "unitary" transformation
\begin{eqnarray}
\eul^{\im\vartheta^a_{\mu}\varrho(p_a^{\mu})}\psi(e^a_{\mu})=\psi(e^a_{\mu}+\vartheta^a_{\mu})\nonumber
\end{eqnarray}
maps vectors from $\Hil$ out of this space, therefore the operators 
$\varrho(p_a^{\mu})$ are not selfadjoint (they are neither symmetric). What we can do with that? We know, thanks 
to the Stone's theorem, that every one-parametric strongly continuous unitary group is related to the 
selfadjoint
operator and vice versa. This implies that if we wish to find the selfadjoint operators for the momenta 
or their functions, we need to find certain groups acting on the space $\conf$. Indeed, a following 
statement is valid.
%\begin{st}
\\\\
Let $\XXX\subset\R^n$ and $\dd x$ be the Lebesgue measure on $\R^n$. If $U(t)$ is one-parametric unitary group
acting on the Hilbert space $\Hil=\Lfun{2}{\XXX}{g\dd x}$, where $g\ge 0$ is locally integrable function on 
$\XXX$, and if $\Phi_t$ is a continuous flow on $\XXX$ 
associated with $U(t)$, then $U(t)$ is strongly continuous.
%\end{st}
\\\\
A proof of the statement is based on the fact that function $I(t):\R\to\R$, defined as
\begin{eqnarray}
I(t)=\int\limits_{\Phi_t^*(K)}f\dd x,\nonumber
\end{eqnarray}
is continuous, where $\Phi_t:\XXX\times\R\to\XXX$ is continous mapping, $K$ is compact subset of $\XXX$ and
$f$ is locally integrable function. It is sufficient to prove that $\|(1-U(t))\psi\|$ is continuous in $t=0$
for all $\psi\in\mathfrak{D}$, where $\mathfrak{D}$ is some dense subset in $\Lfun{2}{\XXX}{g\dd x}$,
since for any convergent sequence $\psi_n\in\mathfrak{D}\to\psi_0\in\Lfun{2}{\XXX}{g\dd x}$ we have
\begin{eqnarray}
\|(1-U(t))\psi_0\|\le\|(1-U(t))(\psi_0-\psi_n)\|+\|(1-U(t))\psi_n\|\le 2\|\psi_0-\psi_n\|
+\|(1-U(t))\psi_n\|.\nonumber
\end{eqnarray}
The set of simple functions is dense in $\Lfun{2}{\XXX}{g\dd x}$, hence for the general simple function
\begin{eqnarray}
f=\sum^m_{i=1}f_i\chiup_{K_i},\nonumber
\end{eqnarray}
where $m\in\N$, $f_i$ are complex constants, $K_i\subset\XXX$ are compacts and 
$K^o_i=K_i\setminus\partial K_i$ are mutually disjoint, 
we have
\begin{eqnarray}
\|(1-U(t))f\|^2&=&\sum_{i,j=1}^m\int g\,\dd x\,\bar{f_i}f_j
\left(\chiup_{K_i}\chiup_{K_j}+\chiup_{\Phi^*_t(K_i)}\chiup_{\Phi^*_t(K_j)}-
\chiup_{\Phi^*_t(K_i)}\chiup_{K_j}-\chiup_{K_i}\chiup_{\Phi^*_t(K_j)}
\right)\nonumber\\
&=&\sum_{i=1}^m\int g\,\dd x\,|f_i|^2
\left(\chiup_{K_i}+\chiup_{\Phi^*_t(K_i)}\right)-
\sum_{i,j=1}^n\int g\,\dd x\,\bar{f_i}f_j
\left(
\chiup_{\Phi^*_t(K_i)\cap K_j}+\chiup_{K_i\cap\Phi^*_t(K_j)}
\right),
\nonumber
\end{eqnarray}
what is continuous in $t$. Hence $U(t)$ is strongly continuous.\\
\hs Now we can try to find group(s) acting on the space $\conf$. The positive linear group $\GL^+$
is not a good candidate, since, as before in the case of $p^{\mu}_a$, there exists transformation $g$ 
from $\GL^+$ which does not preserve the space $\conf$, e.g. rotation in a plane spaned on $e^0_t$, $e^1_t$
maps $e^0_t\to-e^0_t$ and $e^1_t\to-e^1_t$. The problem is caused by the fact that group $\GL^+$ ignores
a metric $\eta_{ab}$. Indeed, if we consider a Lorentz group acting on $e^a_{\mu}$ via
\begin{eqnarray}
e^a_{\mu}\to\left(\eul^{\Lambda\eta}\right)^a_b e^b_{\mu},\label{Lorentz-prva_akcia}
\end{eqnarray}
where $(\Lambda\eta)^a_b=\Lambda^{ac}\eta_{cb}$ and $\Lambda^{ab}=-\Lambda^{ba}$, then we have that
$\eul^{\Lambda\eta}(\conf)\subset\conf$ and even more the transformation (\ref{Lorentz-prva_akcia}) 
is continuous. We can define an operator
\begin{eqnarray}
\UL(\Lambda^{ab})\psi(e^a_{\mu})=\psi\left(\left(\eul^{\Lambda\eta}\right)^a_b e^b_{\mu}\right),
\end{eqnarray}
which is, thanks to the invariance of the measure $\frac{de}{e^4}$, unitary. Let $\Lambda^{ab}$ be arbitrary,
but fixed, then
\begin{eqnarray}
U_{\Lambda}(t)=\UL(t\Lambda^{ab})\nonumber
\end{eqnarray}
is the one-parametric strongly continuous unitary group and, due to the Stone's theorem, we have
that its generator is a selfadjoint operator. We have fixed arbitrary $\Lambda^{ab}$, hence
we have for every $\Lambda^{ab}$ its own generator. $\Lambda^{ab}$ has six degrees of freedom, thus 
there are six independent generators $\Lgen_{ab}$ and we can write
\begin{eqnarray}
\UL(\Lambda^{ab})=\eul^{\im\frac{1}{2}\Lambda^{ab}\Lgen{ab}}\nonumber.
\end{eqnarray}
Let $\psi(e^a_{\mu})\in C^{\infty}_{\text{C}}(\conf)\subset\Hil$, where $C^{\infty}_{\text{C}}(\conf)$ is the
set of all $\infty$-times differentiable functions with compact support on $\conf$, which is dense in $\Hil$, then 
we can use Taylor expansion 
\begin{eqnarray}
\UL(t\Lambda^{ab})\psi(e^a_{\mu})&=&\psi\left(\left(\eul^{t\Lambda\eta}\right)^a_b 
e^b_{\mu}\right)=
\psi\left(
	e^a_{\mu}+
	\left(
					\left(\eul^{t\Lambda\eta}\right)^a_b e^b_{\mu}-e^a_{\mu}
	\right)
\right)=\nonumber\\
&=&\psi(e^a_{\mu})+t\Lambda^{ac}\eta_{cb}e^b_{\mu}\partial_{e^a_{\mu}}\psi(e^a_{\mu})+
t^2o(t, e^a_{\mu})
\label{taylor_exp_Lgen}
\end{eqnarray}
where $o(t, e^a_{\mu})$ is some $C^{\infty}$-function on $\R\times\conf$ with compact support 
on $\conf$ for every $t$ given by Taylor's expansion remainder. The remainder $o(t,e^a_{\mu})$ can 
be restricted for $|t|<\delta$ as $|o(t,e^a_{\mu})|\le M\chiup_{\bar{K}_{\delta}}$, where 
\begin{eqnarray}
K_{\delta}=\cup_{|t|<\delta}K_t,\nonumber
\end{eqnarray}
$K_t$ is a support of $o(t,e^a_{\mu})$ in $\conf$ for given $t$. Since the closure of 
$\cup_{|t|<\delta}\{t\}\times K_t$ is compact in $\R\times\conf$ we have that closure $\bar{K}_{\delta}$
is also compact in $\conf$. Now we can compute the generator $\Lgen(\Lambda^{ab})=
\frac{1}{2}\Lambda^{ab}\Lgen_{ab}$ as a limit $t\to 0$
\begin{eqnarray}
\im\Lgen(\Lambda^{ab})\psi=
\lim_{t\to 0}\frac{\UL(t\Lambda^{ab})-1}{t}\psi.\nonumber
\end{eqnarray}
If we use expansion (\ref{taylor_exp_Lgen}), then we have
\begin{eqnarray}
\frac{1}{t}\left\|\left(\UL(t\Lambda^{ab})-1\right)\psi-\im t\Lgen(\Lambda^{ab})\psi\right\|^2&=&
\frac{1}{t}
\int\limits_{\conf} 
\left|
t\Lambda^{ac}\eta_{cb}e^b_{\mu}\partial_{e^a_{\mu}}\psi+t^2o(t, e^a_{\mu})
		-\im t\Lgen(\Lambda^{ab})\psi
\right|^2\frac{\dd e}{e^4}\le\nonumber\\
&\le&tM^2\int\limits_{K_{\delta}}\frac{\dd e}{e^4},\nonumber
\end{eqnarray}
iff 
\begin{eqnarray}
\Lgen(\Lambda^{ab})=\frac{1}{2}\Lambda^{ab}\Lgen_{ab}=
-\im\Lambda^{ab}\eta_{bc}e^c_{\mu}\partial_{e^a_{\mu}}=
-\im\Lambda^{ab}\eta_{bc}\lambda^{c}\partial_{\lambda^{a}}
-\im\Lambda^{ab}\eta_{bc}E^{c}_{\alpha}\partial_{E^{a}_{\alpha}}.
\label{Lgen}
\end{eqnarray}
Thus we have as a final conclusion that the operator $\Lgen(\Lambda^{ab})$, given by previous expression, 
with domain $\mathfrak{D}(\Lgen(\Lambda^{ab}))=C^{\infty}_{\text{C}}(\conf)$
is essentially selfadjoint for every $\Lambda^{ab}$.\\
\hs This is not everything what the Lorentz group can show us. Let us use again (3+1)-decomposition
$e^a_{\mu}=(\lambda^a,E^a_{\alpha})$. As we already know $\lambda^a$ are components of vector $\partial_t$
in the frame $\ee_a$. Since the time vector can be choosen arbitrary there is no reason to have tied variables $\lambda^a$, $E^a_{\alpha}$ together. Hence we can work with $\lambda^a$, $E^a_{\alpha}$ independently. Let us consider Lorentz group acting on $\lambda^a$ via flow
\begin{eqnarray}
\Phi^{\lambdaup}_t(\Lambda):(\lambda^a,\EE^a)\to
\left(\overline{\lambda}^a,\overline{\EE}^a\right)=\left((\eul^{t\Lambda\eta})^a_b\lambda^b,\EE^a\right),
\end{eqnarray}
then the unitary group corresponding to this flow is
\begin{eqnarray}
\Uni^{\lambdaup}_t(\Lambda):\psiup(\lambda,\EE)\to\psiup(\eul^{t\Lambda\eta}\lambda,\EE)\frac{e^2}{\bar{e}^2},
\end{eqnarray}
where $\bar{e}$ is a determinant of transformed variables $\left(\overline{\lambda}^a,\overline{\EE}^a\right)$. The corresponding selfadjoint \nobreak{generator}
is given by
\begin{eqnarray}
\Llgen(\Lambda)=-\im\Lambda^{ab}\eta_{bc}\lambda^c\partial_{\lambda^a}+
2\im\Lambda^{ab}\eta_{bc}\lambda_a\lambda^c.
\end{eqnarray}
The Lorentz action on $E^a_{\alpha}$ via flow
\begin{eqnarray}
\Phi^{\EE}_t(\Lambda):(\lambda^a,\EE^a)\to\left(\overline{\lambda}^a,\overline{\EE}^a\right)=
\left(\lambda^a,(\eul^{t\Lambda\eta})^a_b\EE^b\right)
\end{eqnarray}
gives unitary group
\begin{eqnarray}
\Uni^{\EE}_t(\Lambda):\psiup(\lambda,\EE)\to\psiup(\lambda,\eul^{t\Lambda\eta}\EE)\frac{e^2}{\bar{e}^2},
\end{eqnarray}
where $\bar{e}$ is again a determinant of transformed variables. The generator is
\begin{eqnarray}
\LEgen(\Lambda)=-\im\Lambda^{ab}\eta_{bc}E_{\alpha}^c\partial_{E^a_{\alpha}}+
2\im\Lambda^{ab}\eta_{bc}\EE_a^c.
\end{eqnarray}
Let us compare these results with (\ref{Lgen}), we can see that
\begin{eqnarray}
\Lgen_{ab}=\Llgen_{ab}+\LEgen_{ab}\nonumber
\end{eqnarray}
as one expected. Generators $\Llgen_{ab}$, $\LEgen_{ab}$ play an important role, since, as we 
will see in a while, their classical 
analogues can be used as coordinates on the phase space.\\
\hs Lorentz group does not change lengths of the vectors, while $\partial_t$ can be 
arbitrary long. We need to cover this feature of $\partial_t$. Let us define a following transformation
\begin{eqnarray}
\lambda^a\to \eul^N\lambda^a,\text{\hs} E^a_{\alpha}\to E^a_{\alpha}.\nonumber
\end{eqnarray}
Let $\Upi	(N)$ be its unitary operator defined via
\begin{eqnarray}
\Upi(N)\psi(\lambda^{a},E^a_{\alpha})=\psi(\eul^N\lambda^{a},E^a_{\alpha})\nonumber
\end{eqnarray}
and its selfadjoint generator is
\begin{eqnarray}
\piup=-\im\lambda^{a}\partial_{\lambda^a}\label{Pigen}.
\end{eqnarray}
A final transformation acting on the space $\conf$ is given by group $\GL^+(\R^3)\equiv\GLs$
acting on the spatial indices $\alpha$. Let $\thetaup^{\alpha}_{\beta}$ be an arbitrary real matrix, 
then the transformation given by
\begin{eqnarray}
\lambda^a\to\lambda^a,\text{\hs} E^a_{\alpha}\to\left(\eul^{\thetaup}\right)_{\alpha}^{\beta}E^a_{\beta}
\end{eqnarray}
represents the change of spatial frame $\partial_{\alpha}\to(\eul^{\thetaup})_{\alpha}^{\beta}\partial_{\beta}$.
Since the transformation does not change a signature of $q_{\alpha\beta}=\eta_{ab}E^a_{\alpha}E^b_{\beta}$,
we have that $\eul^{\thetaup}\conf\subset\conf$ and operators
\begin{eqnarray}
\Uni^{\Delta}(\thetaup)\psi(\lambda^a, E^a_{\alpha})=
\psi\left(\lambda^a,\left(\eul^{\thetaup}\right)_{\alpha}^{\beta}E^a_{\beta}\right)\nonumber
\end{eqnarray}
are unitary and their selfadjoint generators are
\begin{eqnarray}
\Delta^{\alpha}_{\beta}=-\im E^a_{\beta}\partial_{E^a_{\alpha}}.\nonumber
\end{eqnarray}
Let us summarize our situation. We have constructed family of unitary transformations with
actions in the space $\conf$. Now it is a time to find classical variables associated with their
generators. Let us focus on the last four families of the generators.
We have
\begin{eqnarray}
\LLlgen(\Lambda)&=&\Lambda^{ab}\eta_{bc}\lambda^c\pi_a,\nonumber\\
\pi(N)&=&N\lambda^a\pi_a,\nonumber\\
\LLEgen(\Lambda)&=&\Lambda^{ab}\eta_{bc}E^c_{\alpha}G^{\alpha}_a,\nonumber\\
\bldel(\thetaup)&=&\thetaup^{\beta}_{\alpha}E^a_{\beta}G^{\alpha}_a.\nonumber
\end{eqnarray}
Quantum commutators and their classical analogues are
\begin{eqnarray}
 \left[\lambda(k),\piup(N)\right]=\im\lambda(Nk)
&\leftrightarrow&
\left\{\lambda(k),\blpi(N)\right\}=\lambda(Nk)
\nonumber\\
 \left[\lambda(k),\Llgen(\Lambda)\right]=\im\lambda(k\Lambda\eta)
&\leftrightarrow&
\left\{\lambda(k),\LLlgen(\Lambda)\right\}=\lambda(k\Lambda\eta)
,\nonumber\\
 \left[ \Llgen (\Lambda), \Llgen (\Lambda ') \right]=-\im \Llgen(\Lambda\eta\Lambda '-\Lambda '\eta\Lambda)
&\leftrightarrow&
\{\LLlgen(\Lambda),\LLlgen(\Lambda' )\}=-\LLlgen(\Lambda\eta\Lambda '-\Lambda '\eta\Lambda)
,\nonumber\\
\left[\EE(\hh),\Delta(\thetaup)\right]=\im\EE(\thetaup(\hh))
&\leftrightarrow&
\{\EE(\hh),\bldel(\thetaup)\}=\EE(\thetaup(\hh))
,\nonumber\\
\left[\EE(\hh),\LEgen(\Lambda)\right]=\im\EE(\Lambda\eta\hh)
&\leftrightarrow&
\{\EE(\hh),\LLEgen(\Lambda)\}=\EE(\Lambda\eta\hh)
,\nonumber\\
\left[\LEgen(\Lambda),\LEgen(\Lambda ')\right]=\im\LEgen(\Lambda\eta\Lambda '- \Lambda '\eta\Lambda)
&\leftrightarrow&
\{\LLEgen(\Lambda),\LLEgen(\Lambda ')\}=-\LLEgen(\Lambda\eta\Lambda '- \Lambda '\eta\Lambda)
,\nonumber\\
\left[\Delta(\thetaup),\Delta(\thetaup ')\right]=-\im\Delta(\thetaup\thetaup '-\thetaup '\thetaup)
&\leftrightarrow&
\{\bldel(\thetaup),\bldel(\thetaup ')\}=-\bldel(\thetaup\thetaup '-\thetaup '\thetaup)
,\nonumber
\end{eqnarray}
where $\lambda(k)=k_a\lambda^a$, $\EE(\hh)=h^{\alpha}_aE^a_{\alpha}$ and $\Big(\thetaup(\hh)\Big)^{\alpha}_a=\thetaup^{\alpha}_{\beta}h^{\beta}_a$.\\
\hs As we can see we have constructed a selfadjoint representation of the variables
on the space $\Hil=\Lfun{2}{\conf}{\frac{\dd e}{e^4}}$. The question is whether these variables
seperate points of the phase space. Now, we will show that the answer is affirmative. The variables
$\lambda^a$, $\EE^a$ are clear, so let us turn our attention on $\LLlgen_{ab}$, $\blpi$, 
$\LLEgen_{ab}$, $\bldel^{\alpha}_{\beta}$. We have
\begin{eqnarray}
\LLlgen_{ab}\lambda^aE^b_{\alpha}&=&-(\lambda)^2\pi_aE^a_{\alpha}+
\lambda^a\pi_a\eta_{bc}\lambda^cE^c_{\alpha},\nonumber\\
\pi&=&\pi_a\lambda^a,\nonumber\\
\LLEgen_{ab}\lambda^aE^b_{\alpha}&=&q_{\alpha\beta}G^{\beta}_a\lambda^a-
\eta_{ab}\lambda^aE^b_{\beta}G^{\beta}_cE^b_{\alpha},\nonumber\\
\Delta^{\alpha}_{\beta}&=&E^a_{\beta}G^{\alpha}_a,\nonumber
\end{eqnarray}
where $(\lambda)^2=\eta_{ab}\lambda^a\lambda^b$. As we can see, we can invert these equations and we can
express canonical momenta $\pi_a$, $G^{\alpha}_a$ as functions of new variables. The projected variables
$\LLlgen_{\bar{a}\bar{b}}\EE^{\bar{a}}_a\EE^{\bar{b}}_b$, 
$\LLEgen_{\bar{a}\bar{b}}\EE^{\bar{a}}_a\EE^{\bar{b}}_b$ are not independent. They play similar roles like
angular momenta in quantum mechanincs. So, we have found representation of algebra of new variables.
%%%%%%%%%%%%%%%%%%%%%%%%%%%%%%%%%%%%%%%%%%%%%%%%%%%%%%%%%%%%%%%%%%%%%%%%%%%%%%%%%%%%%%%%%%%%%%%%%%%%%%%%%%

\section{Tensor Product Hilbert Space}\label{Tensor_prd}
In the previous section we have constructed the Hilbert space $\Hil_{\xxx}$ associated
with the point $\xxx\in\Sigma$ as $\Hil_{\xxx}=\Lfun{2}{\conf_{\xxx}}{\ems_{\xxx}}$, where 
$\ems=\frac{\dd e}{e^4}$ and $\xxx$ means that it is taken at the point $\xxx$. A main goal
of this section is to briefly summarize ideas of von Neumann's article on tensor product
of family of Hilbert spaces labeled by index set of arbitrary cardinality (details can be found 
in \cite{vNe}). In our case we can formally write
\begin{eqnarray}
\Hil_{\Sigma}=\otimes_{\xxx\in\Sigma}\Hil_{\xxx}.\nonumber
\end{eqnarray}
We have a set $\{\Hil_{\xxx}\}_{\xxx\in\Sigma}$ of Hilbert spaces's labeled by points of $\Sigma$.
A sequence of the states $\{\psi_{\xxx}\}_{\xxx\in\Sigma}$ belongs to the Cartesian product 
$\Hil^{\times}_{\Sigma}=\times_{\xxx\in\Sigma}\Hil_{\xxx}$, but this space is too large, we need to 
pick up a certain subset of $\Hil^{\times}_{\Sigma}$. Let us call $\{\psi_{\xxx}\}_{\xxx\in\Sigma}$
a $C$-sequence iff a product 
\begin{eqnarray}
\|\{\psi_{\xxx}\}_{\xxx\in\Sigma}\|=\prod_{\xxx\in\Sigma}\|\psi_{\xxx}\|_{\xxx}\label{produkt}
\end{eqnarray}
converges. Let $C_{\Sigma}=\{\{\psi_{\xxx}\}_{\xxx\in\Sigma}\text{: $C$-sequence}\}$ 
be a set of all $C$-sequences. A value of the product limit (\ref{produkt}) can be positive or zero.
We need some criteria for convergence of such limits. They can be found in (\cite{vNe}).\\
\\
\textbf{Citation}($\alpha$ - index and $I$ is an index set with arbitrary cardinality):\\
\\
\textit{Lemma} 2.4.1.(p.13):\\
\\
If all $z_{\alpha}$ are real and $\ge 0$, then\\
$\phantom{I}$(I) $\prod_{\alpha\in I}z_{\alpha}$ converges if and only if either 
$\sum_{\alpha\in I}\text{Max}(z_{\alpha}-1,0)$ converges, or some $z_{\alpha}=0$\\
(II) $\prod_{\alpha\in I}z_{\alpha}$ converges and is $\neq 0$ if and only if 
$\sum_{\alpha\in I}|z_{\alpha}-1|$ converges and all $z_{\alpha}\neq 0$.\\
\\
\textit{Lemma} 2.4.2.(p.15):\\
\\
If the $z_{\alpha}$ are arbitrary complex numbers, then $\prod z_{\alpha}$ converges
if and only if\\
$\phantom{I}$(I) either $\prod_{\alpha\in I}|z_{\alpha}|$ converges and its value is $0$,\\
(II) or $\prod_{\alpha\in I}|z_{\alpha}|$ converges and its value is $\neq 0$, and
$\sum_{\alpha\in I}|\text{arcus}\,z_{\alpha}|$ converges\footnote{Is $z\neq 0$, $z=|z|\eul^{\theta}$ with 
$-\pi<\theta\le\pi$, then $\text{arcus}\,z=\theta$}\\
\\
\textit{Definition} 2.5.1.(p.18):\\
\\
$\prod_{\alpha\in I}z_{\alpha}$ is quasi-convergent if and only if $\prod_{\alpha\in I}|z_{\alpha}|$ is
convergent. Its value is\\
\phantom{I}(I) the value of $\prod_{\alpha\in I}z_{\alpha}$ if it is even convergent\\
(II) 0, if it is not convergent.\\
\\
\textbf{End of citation}.\\
\\
\hs The reason why we need a notion of quasi-convergence is that if 
$\{\psi_{\xxx}\}_{\xxx\in\Sigma},\{\phi_{\xxx}\}_{\xxx\in\Sigma}\in C_{\Sigma}$ then product 
$\prod_{\xxx\in\Sigma}\langle \psi_{\xxx}|\pi_{\xxx}\rangle_{\xxx}$ is only quasi-convergent 
in general.\\
Now we can define a functional $\psi_{\Sigma}$ associated with $\{\psi_{\xxx}\}_{\xxx\in\Sigma}$ on
the set $C_{\Sigma}$ of all $C$-sequences as
\begin{eqnarray}
\psi_{\Sigma}(\{\phi_{\xxx}\}_{\xxx\in\Sigma})=\prod\limits_{\xxx\in\Sigma}\langle\phi_{\xxx}|
\psi_{\xxx}\rangle_{\xxx}, \nonumber
\end{eqnarray}
where $\{\phi_{\xxx}\}_{\xxx\in\Sigma}\in C_{\Sigma}$ and product is taken in the sence of
quasi-convergence. It should be noted that 
$\psi_{\Sigma}=0$ does not imply that $\{\psi_{\xxx}=0\}_{\xxx\in\Sigma}$, e.g. for
$C$-sequence $\{\psi_{\xxx_0}=0,\{\psi_{\xxx}\}_{\xxx\in\Sigma\setminus\{\xxx_0\}}\}$ its associated
functional vanishes on whole $C_{\Sigma}$. Let us define a complex linear space $\Hil^0_{\Sigma}$ of
such functionals, where 
\begin{eqnarray}
(a\psi_{\Sigma}+b\phi_{\Sigma})(\{\omega_{\xxx}\}_{\xxx\in\Sigma})=
a^*\psi_{\Sigma}(\{\omega_{\xxx}\}_{\xxx\in\Sigma})+b^*\phi_{\Sigma}(\{\omega_{\xxx}\}_{\xxx\in\Sigma})\nonumber.
\end{eqnarray}
We can define an inner product on $\Hil^0_{\Sigma}$ as follows
\begin{eqnarray}
\langle\psi_{\Sigma}|\phi_{\Sigma}\rangle=\prod\limits_{\xxx\in\Sigma}
\langle\psi_{\xxx}|\phi_{\xxx}\rangle_{\xxx}.\label{inprd}
\end{eqnarray}
The closure $\Hil_{\Sigma}=\overline{\Hil^0_{\Sigma}}$ in the topology defined via inner product (\ref{inprd})
is a Hilbert space and we call it as a tensor product of the sequence $\{\Hil_{\xxx}\}_{\xxx\in\Sigma}$
\begin{eqnarray}
\Hil_{\Sigma}=\otimes_{\xxx\in\Sigma}\Hil_{\xxx}.
\end{eqnarray}
We wish to characterize the space $\Hil_{\Sigma}$ is some way. In order to do so we need to introduce
a notion of $C_0$-sequence and classes of equivalence on them. A sequence $\{\psi_{\xxx}\}_{\xxx\in\Sigma}$
is a $C_0$-sequence iff $\sum_{\xxx\in\Sigma}\big{|}\|\psi_{\xxx}\|_{\xxx}-1\big{|}$ converges. Every $C_0$-sequence is a 
$C$-sequence and every $C$-sequence $\{\psi_{\xxx}\}_{\xxx\in\Sigma}$ is a $C_0$-sequence iff its functional
$\psi_{\Sigma}\neq 0$. We will say that two $C_0$-sequences  are equivalent
$\{\psi_{\xxx}\}_{\xxx\in\Sigma}\sim\{\phi_{\xxx}\}_{\xxx\in\Sigma}$  iff 
$\sum_{\xxx\in\Sigma}\big{|}\langle\psi_{\xxx}|\phi_{\xxx}\rangle_{\xxx}-1\big{|}$ converges, what
is equivalent to the mutual convergence of both series $\sum_{\xxx\in\Sigma}\|\psi_{\xxx}-\phi_{\xxx}\|^2$,
$\sum_{\xxx\in\Sigma}\big{|}\Im(\langle\psi_{\xxx}|\phi_{\xxx}\rangle_{\xxx})\big{|}$, where $\Im(z)$ is
the imaginary part of $z$. Hence we see immediately that if 
$\{\psi_{\xxx}\}_{\xxx\in\Sigma}$, $\{\phi_{\xxx}\}_{\xxx\in\Sigma}$ differ
in finite number of points of $\Sigma$ then they are equivalent. Let us label equivalence classes by $\gamma$ 
and a set of all equivalence classes on $\Hil_{\Sigma}$ by $\EQC$.\\
\hs Now we can finish this bries summary of \cite{vNe} with the following statement. If two $C_0$-sequences
$\{\psi_{\xxx}\}_{\xxx\in\Sigma}$, $\{\phi_{\xxx}\}_{\xxx\in\Sigma}$ or their functional $\psi_{\Sigma}$, 
$\phi_{\Sigma}$ belong to two equivalence classes $\gamma(\psi_{\Sigma})\neq\gamma(\phi_{\Sigma})$, then
$\langle\psi_{\Sigma}|\phi_{\Sigma}\rangle=0$.
If $\gamma(\psi_{\Sigma})=\gamma(\phi_{\Sigma})$ and $\langle\psi_{\Sigma}|\phi_{\Sigma}\rangle=0$ then there
exists $\xxx_0$ where $\langle\psi_{\xxx_0}|\phi_{\xxx_0}\rangle_{\xxx_0}=0$. Hence we see that 
$\Hil_{\Sigma}$ can be decomposed as 
\begin{eqnarray}
\Hil_{\Sigma}=\oplus_{\gamma\in\EQC}\Hil_{\gamma},
\end{eqnarray}
where $\Hil_{\gamma}$ is a Hilbert space associated with $\gamma$.\\
\hs We will use a following example later. Let $K_{\Sigma}=\{K_{\xxx}\}_{\xxx\in\Sigma}$ be sequence of compact sets 
where $K_{\xxx}\subset\conf_{\xxx}$. $K_{\Sigma}$ can be identified with Cartesian product 
$\times_{\xxx\in\Sigma}K_{\xxx}$. Let us define a sets of all sequences of compact sets with unit measure 
as
\begin{eqnarray}
J^1(\Conf)=\Big{\{}K_{\Sigma}=\{K_{\xxx}\}_{\xxx\in\Sigma}:\forall \xxx\in\Sigma;\,\,\,\, \ems_{\xxx}(K_{\xxx})=1\Big{\}}\nonumber
\end{eqnarray}
We can associate with $K_{\Sigma}\in J^1(\Conf)$ an element in $\Hil_{\Sigma}$ via
\begin{eqnarray}
\chiup_{K_{\Sigma}}=\{\chiup_{K_{\xxx}}\}_{\xxx\in\Sigma}.
\end{eqnarray}
Let $K_{\Sigma},K'_{\Sigma}\in J^1(\Conf)$ and $\sigma\subset\Sigma$ be a set of all $\xxx$ where 
$K_{\xxx}\neq K'_{\xxx}$. We will use a notation $\ee=(e^a_{\mu})$, $\ee_{\xxx}=(e^a_{\mu}\vb{\xxx})$.
Let $\ee\in K_{\Sigma}\setminus K'_{\Sigma}$. If we suppose that for $\forall\xxx\in\sigma$ exists
an open neighbourhood of $\ee_{\xxx}\in U_{\xxx}$ with property 
$\overline{U_{\xxx}}\subset K_{\xxx}\setminus K'_{\xxx}$ and $\ems_{\xxx}(U_{\xxx})>\delta\in (0,1)$ and 
$\sigma$ is not a finite set then 
\begin{eqnarray}
\langle\chiup_{K_{\Sigma}}|\chiup_{K'_{\Sigma}}\rangle=\prod_{\xxx\in\Sigma}\ems_{\xxx}
(K_{\xxx}\cap K'_{\xxx})=0,\nonumber
\end{eqnarray}
since $1>1-\delta>1-\ems_{\xxx}(K_{\xxx}\setminus K'_{\xxx})=\ems_{\xxx}(K_{\xxx}\cap K'_{\xxx})$.

%%%%%%%%%%%%%%%%%%%%%%%%%%%%%%%%%%%%%%%%%%%%%%%%%%%%%%%%%%%%%%%%%%%%%%%%%%%%%%%%%%%%%%%%%%%%%%%%%%%%%%%%%%
\section{Quantum Algebra of Basic Variables}\label{Quantum_algebra}
Now it is time to construct a representation of the basic variables of the Einstein-Cartan theory. Inspired
by the point version of the phase space we will not work with canonical variables, but we will construct a 
representation of the following variables
\begin{eqnarray}
\lambda(\kk)&=&\ints\kk_a\lambda^a,\nonumber\\
\LLlgen(\Lambda)&=&\int\limits_{\Sigma}\Lambda^{ab}\eta_{bc}\lambda^c\blpi_a,\nonumber\\
\blpi(N)&=&\int\limits_{\Sigma}N\lambda^a\blpi_a,\nonumber\\
\EE(\HH)&=&\ints\HH_a\wedge\EE^a,\nonumber\\
\LLEgen(\Lambda)&=&\int\limits_{\Sigma}\Lambda^{ab}\eta_{bc}\EE^c\wedge\GGG_a,\nonumber\\
\bldel(\thetaup)&=&\int\limits_{\Sigma}\thetaup(\EE^a)\wedge\GGG_a.\nonumber
\end{eqnarray}
where $\thetaup(\EE^a)=E^a_{\alpha}\thetaup^{\alpha}_{\beta}\dd x^{\beta}$, with similar algebra as in the point version (trivial brackets are not written)
\begin{eqnarray}
 &\Big{\{}\lambda(\kk)\,,\,\blpi(N)\,\Big{\}}^*&=\,\,\,\phantom{-}\lambda(N\kk),
\nonumber\\
 &\Big{\{}\lambda(\kk),\LLlgen(\Lambda)\Big{\}}^*&=\,\,\,\phantom{-}\lambda(\kk\Lambda\eta),
\nonumber\\
 &\Big{\{}\LLlgen(\Lambda),\LLlgen(\Lambda' )\Big{\}}^*&=\,\,\,-\LLlgen(\Lambda\eta\Lambda '-\Lambda '\eta\Lambda),
\nonumber\\
&\Big{\{}\EE(\HH),\bldel(\thetaup)\Big{\}}^*&=\,\,\,\phantom{-}\EE(\thetaup(\HH)),
\nonumber\\
&\Big{\{}\EE(\HH),\LLEgen(\Lambda)\Big{\}}^*&=\,\,\,\phantom{-}\EE(\Lambda\eta\HH),
\nonumber\\
&\Big{\{}\LLEgen(\Lambda),\LLEgen(\Lambda ')\Big{\}}^*&=\,\,\,-\LLEgen(\Lambda\eta\Lambda '- \Lambda '\eta\Lambda),\nonumber\\
&\Big{\{}\bldel(\thetaup),\bldel(\thetaup ')\Big{\}}^*&=\,\,\,-\bldel(\thetaup\thetaup '-\thetaup '\thetaup)\nonumber
\end{eqnarray}
\hs Before we start to costruct a representation of this algebra, we need to discuss  properties of a certain
family of operators. Let $\Apx$ be a selfadjoint operator with action on $\Hil_{\xxx}$ with dense domain 
$\mathfrak{D}(\Apx)$. We wish to represent it on the space $\Hil_{\Sigma}$. Since 
$\Hil_{\Sigma}\simeq\Hil_{\xxx}\otimes\Hil_{\Sigma\setminus\{\xxx\}}$ we can see that expression 
\begin{eqnarray}
\Uni_{\Sigma}(t)\psi_{\Sigma}=\big{\{}\Uni_{\xxx}(t)\psi_{\xxx};\{\psi_{\y}\}_{\y\neq\xxx}\big{\}},
\end{eqnarray}
where $\psi_{\Sigma}$ is $C$-sequence, defines an unitary operator on whole $\Hil_{\Sigma}$, 
which is strongly continuous at $t$. 
$\Uni_{\Sigma}(t)$ determines a generator $\Aps$ associated with it  
and $\mathfrak{D}(\Aps)\supset\mathfrak{D}_o(\Apx)=
\text{Span}
\{
\psi_{\xxx}\otimes\psi_{\Sigma\setminus\{\xxx\}};
\psi_{\xxx}\in\mathfrak{D}(\Apx),\psi_{\Sigma\setminus\{\xxx\}}\in\Hil^0_{\Sigma\setminus\{\xxx\}}
\}$. Restricted operator $\Aps\vb{{\mathfrak{D}_o}(\Apx)}$ is essentially selfadjoint and acts on $C$-sequences
$\psi_{\Sigma}\in\mathfrak{D}_o(\Apx)$ as
\begin{eqnarray}
\Aps\vb{\mathfrak{D}_o(\Apx)}\psi_{\Sigma}=\big{\{}\Apx\psi_{\xxx},\{\psi_{\y}\}_{\Sigma\setminus\{\xxx\}}\big{\}}.
\nonumber
\end{eqnarray}
\hs Let us start with variables $\lambda_{\xxx}(k)=\lambda^a(\xxx)k_a(\xxx)$, $\EE_{\xxx}(\hh)=E^a_{\alpha}(\xxx)h^{\alpha}_a(\xxx)$. Both of them are acting on the space
$\Hil_{\xxx}$, hence we can represent them via previous construction on the space $\Hil_{\Sigma}$ by 
formula for $C$-sequence $\psi_{\Sigma}\in C_{\Sigma}$
\begin{eqnarray}
\varrho\Big{(}\lambda_{\xxx}(k)\Big{)}\psi_{\Sigma}=
\Big{\{}
\lambda_{\xxx}(k)\psi_{\xxx}(\ee);\{\psi_{\y}\}_{\y\in\Sigma\setminus\{\xxx\}}
\Big{\}},
\nonumber\\
\varrho\Big{(}\EE_{\xxx}(\hh)\Big{)}\psi_{\Sigma}=
\Big{\{}
\EE_{\xxx}(\hh)\psi_{\xxx}(\ee);\{\psi_{\y}\}_{\y\in\Sigma\setminus\{\xxx\}}
\Big{\}}.\nonumber
\end{eqnarray}
\hs We have used the actions of the groups $\SO^{+}(\eta)$ for $\lambda^a$, $\SO^{+}(\eta)$ for $\EE^a$, $\R^+$ 
and $\GLs$ on the space $\conf$. Now, we wish to generalize this idea to Einstein-Cartan theory.
Let $\GGG_{\xxx}$ be one, same for all $\xxx$, of the previous groups acting on the space $\conf_{\xxx}$ and 
let $\Phi^{\xxx}_t$ be flow associated with some one parametric subgroup of $\GGG_{\xxx}$. Then we have a group $\GGG^{\Sigma}=\times_{\xxx\in\Sigma}\GGG_{\xxx}$ acting on the 
space $\Conf=\times_{\xxx\in\Sigma}\conf_{\xxx}$ by the flow $\Phi^{\Sigma}_t(\ee)=
\{\Phi^{\xxx}_t(\ee_{\xxx})\}_{\xxx\in\Sigma}$. Since for $\forall\xxx\in\Sigma$ the flow $\Phi^{\xxx}_t$ defines unitary operator $\Uni^{\xxx}$ on $\Hil_{\xxx}$ we can define unitary operator associated with 
flow $\Phi^{\Sigma}_t$. Let $\psi^j_{\Sigma}$ ($j=1,\ldots,m\in\N$) be $C$-sequences, then an operator defined for any $\Psi\in\Hil^0_{\Sigma}$
\begin{eqnarray}
\Uni^{\Sigma}(t)\Psi=\sum_{j=1}^{m\in\N}c_{j}\Uni^{\Sigma}(t)\psi^j_{\Sigma}=
\sum_{j=1}^{m\in\N}c_{j}\Big{\{}\Uni^{\xxx}(t)\psi^j_{\xxx}\Big{\}}_{\xxx\in\Sigma},\nonumber
\end{eqnarray}
where $\Psi=\sum_{j=1}^{m\in\N}c_{j}\psi_{\Sigma}^{j}$, 
can be extended to the one-parametric unitary grup acting on whole $\Hil_{\Sigma}$. We know nothing about its
continuity at the moment.\\
\hs Let $K_{\Sigma}\in J^1(\Conf)$ be a constant sequence of compact sets, i.e. 
$\forall\xxx$ $K_{\xxx}=K$, and let $\Phi^{\xxx}_t=\Phi_t$ for $\forall\xxx\in\sigma\subset\Sigma$ and
$\Phi^{\xxx}_t=\text{id}$ for $\forall\xxx\in\Sigma\setminus\sigma$. Let us explore an expression
\begin{eqnarray}
u(t)=\Big{\|}\left(1-\Uni^{\Sigma}_t\right)\chiup_{K_{\Sigma}}\Big{\|}^2.\nonumber
\end{eqnarray}
It is clear by definition, that $u(0)=0$. Let $t\neq 0$, then we can write
\begin{eqnarray}
u(t)=\Big{\langle}\chiup_{K_{\Sigma}}\Big{|}\left(1-\Uni^{\Sigma}_{-t}\right)
\left(1-\Uni^{\Sigma}_t\right)\chiup_{K_{\Sigma}}\Big{\rangle}=
2
-\Big{\langle}\chiup_{K_{\Sigma}}\Big{|}\Uni^{\Sigma}_{-t}\chiup_{K_{\Sigma}}\Big{\rangle}
-\Big{\langle}\chiup_{K_{\Sigma}}\Big{|}\Uni^{\Sigma}_t\chiup_{K_{\Sigma}}\Big{\rangle}.
\nonumber
\end{eqnarray}
The last two terms are zero in the case when $\sigma$ is not finite due to the arguments from
the end of the previous section. Hence we have, as a consequence, that operator $\Uni^{\Sigma}(t)$ is not
strongly continuous in the general case. Therefore there does not exist selfadjoint generator of 
$\Uni^{\Sigma}(t)$  in the general case.\\
\hs What we can do is to explore the case when the group $\GGG^{\Sigma}$ acts on $\Conf$ nontrivially 
only on some finite subset $\sigma\subset\Sigma$. Let us start with $\sigma=\{\xxx\}$. This case were
explored few rows above and point generators $\mathsf{T}_{\xxx}$ of such action were found in 
section 2.2. Generalization to the case when $\sigma=\{\xxx_1,\dots,\xxx_n\}$ is clear and the resulting
generator is $\mathsf{T}_{\sigma}=\sum_{\xxx\in\Sigma}\mathsf{T}_{\xxx}$.\\
\hs Now we can write explicitly the generators of our groups acting on the $\Conf$. They are
\begin{eqnarray}
\piup(N)&=&\sum_{\xxx\in\Sigma}-\im N(\xxx)\lambda^{a}(\xxx)\partial_{\lambda^{a}(\xxx)},\nonumber\\
\Llgen(\Lambda)&=&\sum_{\xxx\in\Sigma}-\im\Lambda^{ab}(\xxx)\eta_{bc}\lambda^c(\xxx)\partial_{\lambda^{a}(\xxx)},
\nonumber\\
\Delta(\thetaup)&=&\sum_{\xxx\in\Sigma}-\im\thetaup^{\beta}_{\alpha}(\xxx)E^a_{\beta}(\xxx)\partial_{E^a_{\alpha}(\xxx)},
\nonumber\\
\LEgen(\Lambda)&=&\sum_{\xxx\in\Sigma}-\im\Lambda^{ab}(\xxx)\eta_{bc}E^b_{\alpha}(\xxx)\partial_{E^a_{\alpha}(\xxx)},
\nonumber
\end{eqnarray}
where $N(\xxx)$, $\Lambda^{ab}(\xxx)$, $\thetaup^{\beta}_{\alpha}(\xxx)$ has support on a finite set. 
Commutator algebra of basic quantum observables is generated by
\begin{eqnarray}
 &\Big{[}\varrho\Big{(}\lambda_{\xxx}(k)\Big{)}\,,\,\piup(N)\,\Big{]}&=
\phantom{-}\im\varrho\Big{(}\lambda_{\xxx}(Nk)\Big{)},
\nonumber\\
&\Big{[}\varrho\Big{(}\lambda_{\xxx}(k)\Big{)},\Llgen(\Lambda)\Big{]}&=
\phantom{-}\im\varrho\Big{(}\lambda_{\xxx}(k\Lambda\eta)\Big{)},
\nonumber\\
 &\Big{[}\Llgen(\Lambda),\Llgen(\Lambda' )\Big{]}&=
-\im\Llgen(\Lambda\eta\Lambda '-\Lambda '\eta\Lambda),
\nonumber\\
&\Big{[}\varrho\Big{(}\EE_{\xxx}(\hh)\Big{)},\Delta(\thetaup)\Big{]}&=
\phantom{-}\im\varrho\Big{(}\EE_{\xxx}(\thetaup(\hh))\Big{)},
\nonumber\\
&\Big{[}\varrho\Big{(}\EE_{\xxx}(\hh)\Big{)},\LEgen(\Lambda)\Big{]}&=
\phantom{-}\im\varrho\Big{(}\EE_{\xxx}(\Lambda\eta\hh)\Big{)},
\nonumber\\
&\Big{[}\LEgen(\Lambda),\LEgen(\Lambda ')\Big{]}&=
-\im\LEgen(\Lambda\eta\Lambda '- \Lambda '\eta\Lambda),\nonumber\\
&\Big{[}\Delta(\thetaup),\Delta(\thetaup ')\Big{]}&=
-\im\Delta(\thetaup\thetaup '-\thetaup '\thetaup).\nonumber
\end{eqnarray}
Hence we see that we found representation of classical variables of Einstein-Cartan theory.\\
\hs Really? This statement needs some additional explanation. The first thing which should be taken into account is given by the fact that on classical level we were working with smooth variables $(\lambda^a(\xxx),E^a_{\alpha}(\xxx))$ as is usual. Since on the quantum level we were looking for representation by von Neumann infinity-dimensional tensor product of point Hilbert spaces where
all points are independend the space $\Conf$ is no longer space of smooth variables but rather the space
of cartesian product $\Conf=\times_{\xxx\in\Sigma}\conf_{\xxx}$. Tangent space $\Td_{\ee}\Conf$ in $\ee\in\Conf$ is isomorphic to $\FFF(\Sigma,\R^{4\times 4})=\Big{(}\R^{4\times 4}\Big{)}^{\Sigma}$ what is the space of all function on $\Sigma$ valued in $\R^{4\times 4}$. Space $\Conf$ and $\FFF(\Sigma,\R^{4\times 4})$ are equipped with standard Tychonov topology. Its topological dual $\FFF^*(\Sigma,\R^{4\times 4})$ is given by linear mappings
\begin{eqnarray}
\alpha(f)=\sum\limits_{\xxx\in\Sigma}\alpha_a^{\mu}(\xxx)f^a_{\mu}(\xxx),\label{dual}
\end{eqnarray}
where $\alpha_a^{\mu}(\xxx)$ has support on finite subset of $\Sigma$. We can write formally instead of (\ref{dual}) a following expression
\begin{eqnarray}
\alpha(f)=\ints\sum_{\xxx ' \in\Sigma}\alpha_a^{\mu}(\xxx ')\delta_{\xxx\xxx '}f^a_{\mu}(\xxx)\dd^3 x=\ints\alpha_a\wedge f^a+\ints\boldsymbol{\alpha}_a\wedge\mathbf{f}^a,\label{dual2}
\end{eqnarray}
where 
\begin{eqnarray*}
&\alpha_a&=\sum_{\xxx ' \in\Sigma}\alpha_a^0(\xxx ')\delta_{\xxx\xxx '}\dd^3x,\\
&\boldsymbol{\alpha}_a&=\sum_{\xxx ' \in\Sigma}\frac{1}{2}\alpha_a^{\alpha}(\xxx ')\delta_{\xxx\xxx '}\varepsilon_{\alpha\beta\gamma} \dd x^{\beta}\wedge \dd x^{\gamma},\\
&f^a&=f^a_0\\
&\mathbf{f}^a&=f^a_{\alpha}\dd x^{\alpha}
\end{eqnarray*}
Since the space $\FFF^*(\Sigma,\R^{4\times 4})$ isomorphic to $\Td_{\ee}^*\Conf$ we can identify
\begin{eqnarray*}
&\blpi_a&=\sum_{\xxx ' \in\Sigma}\pi_a(\xxx ')\delta_{\xxx \xxx '}\dd^3x,\\
&\GGG_a&=\sum_{\xxx ' \in\Sigma}G_a^{\alpha}(\xxx ')\delta_{\xxx\xxx '}\varepsilon_{\alpha\beta\gamma} \dd x^{\beta}\wedge \dd x^{\gamma}.
\end{eqnarray*}
\hs Now, let us explore a reducibility of this representation. As we already know, the space $\Hil_{\Sigma}$
can be decomposed into the mutually orthogonal subspaces labeled by classes of equivalences of $C_0$-sequences $\EQC$. Our representation does not mix this decomposition hence it is reducible. 
Number of irreducible representations in $\Hil$ is equal to the number of equivalence classes on $\Hil_{\Sigma}$, what is "huge" infinite, e.g. for every element of $\Lfun{2}{\conf}{\frac{de}{e^4}}$ there exists its own equivalence class, etc. One may partially save the situation by using unitary version of basic variables and represents operators $\Uni^{\Sigma}_{\GGG_{\Sigma}}(\xiup)$, where e.g. $\xiup=N$ for $\R^+$, etc., instead of its generators $\mathsf{T}(\xiup)$, with action on whole $\Conf$ which mix orthogonal decomposition of  $\Hil_{\Sigma}$. Anyway for $K^1_{\Sigma},K^2_{\Sigma}\in J^1(\Conf)$, where $K^1_{\Sigma}$ is built by simple connected sets and $K^2_{\Sigma}$ is built by union of two simple connected sets, there is no element of  $\GGG_{\Sigma}$ which mixes their equivalence classes and reducibility of unitary representation is still too huge. 
%%%%%%%%%%%%%%%%%%%%%%%%%%%%%%%%%%%%%%%%%%%%%%%%%%%%%%%%%%%%%%%%%%%%%%%%%%%%%%%%%%%%%%%%%%%%%
\section{Comments}\label{comments}
We have defined a point version of the phase space of Einstein-Cartan theory. 
Basic variables of the point phase space were successfully quantized. These results were
used for construction of field variables representation via von Neumann construction of infinity dimensional
tensor product of point Hilbert spaces.\\
\hs The problem of the presented construction lies in the fact, that the canonical momenta $\GGG_a$ does not transform as a tensor under local Lorentz transformation since it is linear funcional of connection potential $\AAA^{ab}$. Dirac bracket between $\GGG_a$ and Lorentz generator $\LLEgen$ generates the tensor type transformation on $\GGG_a$. In other words Lorentz transformation generated by $\LLEgen$ is not physical transformation but rather only the transformation on the absract phase space. The way how to go throught this obstacle is to find a tensorial canonical momentum let say $\PP_a$. In the second paper of the miniseries started by work\cite{Pilc} will be shown know how looks the physical Lorentz generators. They are given by torsion contrains $\T_{ab}$, which can be formally written as $\TT\sim\EE\wedge\dd\EE+\EE\wedge\GGG$.
The basic idea of construction of tensorial momenta is to find it in a form $\PP\sim A\dd\EE+\GGG$, where $A$ depends only on $\lambda, \EE$.\\
\hs Another open problem is given by the question how does look the quantum representation of the first class constraints. This should be explored carefully since our quantum configuration space $\Conf$ is the space of all orthonormal coframes over $\Sigma$ and it does not look that the exterior derivative operator $\dd$ can be defined on such space. Perhaps the Regge ideas can be used here. 
\hs Problem of the huge reducibility is familiar for field theories. If we recall the quantization of scalar (Dirac, Standard model,...) fields then the similar problem was occured there and it was solved by finding the correct vacuum by superselection rule given by condition that the vacuum should be invariant under 
Poincar\' e group. Of course quantum gravity has no Lorentz invariant backround or something like that since we are working with full theory where the geometry of the spacetime is dynamical quantity. But we are already assuming that the final physical Hilbert space is invariant under local Poincar\' e transformations, since the constrains generating such gauge symmetry are driven by EOM to vanish. So the question is whether the condition of vanishing the local Poincar\' e generators will pick up the correct representation subspace of general Hilbert space as it is in the case of usual quantum field theory on Minkowski backround. 

\begin{acknowledgments}
I would like to thank to prof. Ji\v r\' i Bi\v c\' ak for his supervising of my work, to
prof. J. Fernando Barbero G., prof. Martin Bojowald, Dr. Daniele Oriti for their comments to my doctoral thesis. My thanks also belong to Otakar Sv\' itek, Vladim\' ir Balek, Michal Demetrian, Jakub Lehotsk\' y, V\' aclav D\v edi\v c, Richard Richter, Rastislav Mint\' ach for their friendship. 
I am very appreciating the support, love of and life  experience with my Femme Fatale Hana Korbov\' a. Unfortunately I am no more able to thank to my mum thus please 
let me dedicate this work to her, to \v Ludmila Pilcov\' a in memoriam.
\end{acknowledgments}

\bibliography{Bibliography}% Produces the bibliography via BibTeX.

\end{document}